\documentclass[12pt,a4paper]{article}

\usepackage[english]{babel}

\usepackage{indentfirst}
\usepackage{amsmath}
\usepackage{amssymb}
\usepackage[pdftex]{graphicx,color}
\usepackage{cite}
\usepackage{slashed}
\numberwithin{equation}{section}

\usepackage{geometry}
\begin{document}

\hfill {\large IPPP/19/88} \\[5mm]

\begin{center}
\textbf{\Large Probing New Physics in \boldmath $b \to d$ Transitions}

\vspace*{10mm}

{\large Aleksey~V.~Rusov}
 
\vspace*{0.4cm}

\textsl{%
Institute for Particle Physics Phenomenology, Durham University, \\
DH1 3LE Durham, United Kingdom \\[5mm]
}
\vspace*{1.5cm}

\textbf{\large Abstract}\\[10pt]
\parbox[t]{0.9\textwidth}
{Recent experimental data on several observables in semileptonic $B$-meson
decays are found to be in tension with the corresponding Standard Model predictions. 
Most of these deviations are related to $b \to c$ and $b \to s$ 
flavour changing transitions. In this work, we estimate possible 
New Physics effects in $b \to d \mu^+ \mu^-$ flavour changing neutral currents.
We parametrize NP contributions in a model-independent way and determine
the allowed ranges of corresponding Wilson coefficients from the data on the exclusive
$B^\pm \to \pi^\pm \mu^+ \mu^-$ decays measured recently by the LHCb collaboration. 
Afterwards, we investigate the impact of these results on other $b \to d$ processes 
such as the leptonic $B^0 \to \mu^+ \mu^-$ decays and $B^0 - \bar B^0$ mixing.
As an example, we consider a simplified $Z^\prime$ model that is found to be consistent 
with current $b \to d$ data in the certain regions of the NP parameter space.
In addition, we estimate the correlations between the partial decay widths 
of $B \to \pi \mu^+ \mu^-$ and $B \to K \mu^+ \mu^-$ processes to be used 
for an independent determination of CKM matrix elements as well as 
for a combined New Physics analysis of both $b \to d$ and $b \to s$ transitions.}
\vspace*{1cm}

\parbox[t]{0.9\textwidth}
{KEYWORDS: B-physics, Rare decays, New Physics}

\end{center}

\vspace*{3cm}
\pagestyle{empty}
\newpage
\pagestyle{plain}

\section{Introduction}

Current tensions between experimental measurements and Standard Model (SM) predictions 
of some observables in $B$-hadron decays (so-called $B$-anomalies) have attracted a lot of attention. 
Most of these anomalies are related with $b \to c \ell \bar \nu_\ell$ and $b \to s \ell^+ \ell^-$ 
flavour changing transitions where $\ell$ denotes one of the charged leptons. 
To accommodate these tensions, several New Physics (NP) models have been proposed in the literature 
(e.g. with leptoquarks, $Z^\prime$-boson, etc.) leading to the publication of large number of papers
\cite{Khodjamirian:2010vf, Straub:2015ica, Alonso:2015sja, Greljo:2015mma, Barbieri:2015yvd, Calibbi:2015kma, Bauer:2015knc, Crivellin:2015lwa, Fajfer:2015ycq, Bordone:2016gaq, Hiller:2016kry, Bhattacharya:2016mcc, Altmannshofer:2017yso, Bobeth:2017vxj, Capdevila:2017bsm, Geng:2017svp, Ciuchini:2017mik, Hurth:2017hxg,   Ghosh:2017ber, Datta:2017ezo, Altmannshofer:2017bsz, Celis:2017doq, DAmico:2017mtc, Buttazzo:2017ixm, DiLuzio:2017chi, Angelescu:2018tyl, Becirevic:2018afm, Crivellin:2018yvo, Arbey:2018ics, Alguero:2019ptt, Alok:2019ufo, Ciuchini:2019usw, Calibbi:2019lvs, Kowalska:2019ley, Aebischer:2019mlg, Arbey:2019duh, Fuentes-Martin:2019mun, Arnan:2019uhr, Chang:2010vz, Chang:2010gi, Chang:2012zza, Crivellin:2019dwb}.
On the other side, if NP exists and is accessible at current energy level, it would be reasonable 
to expect such effects also in processes induced by $b \to d$ flavour changing neutral current (FCNC). 
Similar to $b \to s \ell^+ \ell^-$ transition, the $b \to d \ell^+ \ell^-$ FCNCs are 
forbidden at tree level in the SM and induced via loops,
therefore they also might be sensitive to NP contributions.
One of the specific features of the $b \to d$ transitions 
is an additional suppression compared to the $b \to s$ by ratio
of the Cabibbo-Kobayashi-Maskawa (CKM) matrix elements $|V_{td}/V_{ts}|^2$.
The typical branching fraction of $b \to d \ell^+ \ell^-$ processes is 
${\cal O} (10^{-8})$ which makes their measurements considerably more challenging. 
Additionally, the parts of the amplitude of the $b \to d \ell^+ \ell^-$ decays 
proportional to $V_{tb} V_{td}^*$, $V_{cb} V_{cd}^*$ and $V_{ub} V_{ud}^*$ 
are of the same order of the Wolfenstein parameter $\lambda$ 
and, in addition to a relative CKM phase, they have different strong phases
originating from the nonlocal hadronic amplitudes. 
This leads to non-vanishing direct $CP$-asymmetry in $b \to d \ell^+ \ell^-$
processes which is negligible in case of $b \to s \ell^+ \ell^-$ transition.
Therefore $b \to d$ processes provide an even richer set of interesting observables
to test the quark flavour sector of the Standard Model.

Up to now, only few semileptonic $b \to d \ell^+ \ell^-$ processes have been seen experimentally. 
The first measurement of the semileptonic $b \to d$ transition was done by the LHCb collaboration
in 2012 providing an experimental value of the branching fraction of the exclusive 
$B^\pm \to \pi^\pm \mu^+ \mu^-$ decays \cite{LHCb:2012de}. 
In 2015, the LHCb collaboration has also measured the differential decay distribution in the dimuon 
invariant mass squared and the total direct $CP$-asymmetry of the $B^\pm \to \pi^\pm \mu^+ \mu^-$
processes~\cite{Aaij:2015nea}.
The LHCb data presented in Ref.~\cite{Aaij:2015nea} is about 1.3$\sigma$
away from the recent SM prediction including the computation of the corresponding 
nonlocal hadronic amplitudes and resonance contributions \cite{Hambrock:2015wka}
(see Fig.~4 in Ref.~\cite{Aaij:2015nea}). Curiously, this slight deviation of experiment and 
theory points in the same direction as the tensions found in the $b \to s \mu^+ \mu^-$ transitions
(as it was also noticed in Ref.~\cite{Crivellin:2018yvo}).
Such a situation, together with current tensions in $b \to c$ and $b \to s$
motivated us to address the main goal of this paper, namely, to probe
possible New Physics effects in $b \to d \ell^+ \ell^-$ processes.

Experimentally also the decays $B^0 \to \pi^+ \pi^- \mu^+ \mu^-$~\cite{Aaij:2014lba}
and $\Lambda_b^0 \to p \pi^- \mu^+ \mu^-$~\cite{Aaij:2017ewm} have been studied. 
The theoretical analysis of these processes is however quite challenging due to a poor knowledge 
of the underlying hadronic input including form factors and non-local hadronic amplitudes, 
therefore we do not include these decays in our NP analysis.
In addition, the LHCb collaboration has recently found evidence of the 
$B_s^0 \to \bar K^{*0} \mu^+ \mu^-$ decay at the level of 3.4 standard deviations \cite{Aaij:2018jhg}.
An analysis of this decay is of special interest in the light of existing anomalies 
in the $B \to K^* \mu^+ \mu^-$ processes.

The paper is organised as follows. In Section~2 we determine 
favored intervals of the NP coefficients in a model-independent way from data on the
differential branching fraction of the $B^\pm \to \pi^\pm \mu^+ \mu^-$ decays.
In Section~3 we consider the impact on the leptonic $B^0 \to \mu^+ \mu^-$ decays.
Section~4 is devoted to an analysis of NP effects in $B^0 - \bar B^0$ mixing:
as an example we study a simplified NP model with $Z^\prime$-boson. 
We conclude in Section~5 and in Appendix we present the correlation matrix 
between different hadronic parts of the partial decays width 
in $B \to \pi \ell^+ \ell^-$ and $B \to K \ell^+ \ell^-$ processes in the SM.

\section{New Physics effects in \boldmath $B^\pm \to \pi^\pm \mu^+ \mu^-$ decays}

We perform an analysis of possible NP effects in the $B^\pm \to \pi^\pm \mu^+ \mu^-$ decays 
in a model-independent way based on an assumption that these effects are induced at a large energy scale 
(by heavy particles e.g. $Z^\prime$-boson, leptoquarks, etc.). 
After integrating out their contributions are described by an effective Lagrangian
\begin{equation}
{\cal L}_{\rm eff}^{\rm NP} = \frac{4 G_F}{\sqrt 2} V_{tb} V_{td}^* \,
\left(C_9^{\mu} {\cal O}_9^{\mu} + C_{10}^{\mu} {\cal O}_{10}^{\mu} 
+ C_9^{\prime \mu} {\cal O}_9^{\prime \mu} + 
C_{10}^{\prime \mu} {\cal O}_{10}^{\prime \mu} \right) + {\rm h.c.}, 
\label{eq:Heff-NP}
\end{equation}
where $G_F$ is the Fermi constant, 
$C_{9,10}^{(\prime)\mu}$ denote the short-distance NP Wilson coefficients,
and the effective dimension-6 semileptonic operators are defined as
\begin{eqnarray}
{\cal O}_9^{\mu} = 
\frac{\alpha_{\rm em}}{4 \pi}
(\bar d \gamma_\rho P_L b) (\bar \mu \gamma^\rho \mu), 
& & \qquad 
{\cal O}_{10}^{\mu} = 
\frac{\alpha_{\rm em}}{4 \pi}
(\bar d \gamma_\rho P_L b) (\bar \mu \gamma^\rho \gamma_5 \mu), \\
{\cal O}_9^{\prime \mu} = 
\frac{\alpha_{\rm em}}{4 \pi}
(\bar d \gamma_\rho P_R b) (\bar \mu \gamma^\rho \mu), & & 
\qquad {\cal O}_{10}^{\prime \mu} =
\frac{\alpha_{\rm em}}{4 \pi} 
(\bar d \gamma_\rho P_R b) (\bar \mu \gamma^\rho \gamma_5 \mu),
\label{eq:NP-operators}
\end{eqnarray}
with $\alpha_{\rm em}$ denoting the fine structure constant, and 
$P_{L, R} = (1 \mp \gamma_5)/2$. 
Here we make several comments regarding the NP ansatz by Eq.~(\ref{eq:Heff-NP}) 
used in our analysis. First, we consider the effective NP operators 
with muons only since the measurements of $b \to d \ell^+ \ell^-$ modes 
with electrons or $\tau$-leptons are absent at present time. Therefore, 
currently Lepton Flavour Universality (LFU) cannot be tested in $b \to d \ell^+ \ell^-$ processes. 
Furthermore, we emphasize that this is just an initial study and 
due to insufficient current data in $b \to d$ transition we restrict ourselves by a simplified ansatz 
in Eq.~(\ref{eq:Heff-NP}) without considering the (pseudo)scalar and tensor as well as electromagnetic,
chromomagnetic and four-quark effective operators. The choice of the NP effective Lagrangian
(\ref{eq:Heff-NP}) is motivated by the $b \to s \ell^+ \ell^-$ case where 
a better agreement with data is achieved from the fit by using the vector NP operators (see e.g. Ref.~\cite{Aebischer:2019mlg}). 

The effective NP Lagrangian (\ref{eq:Heff-NP}) modifies the 
expression for the dilepton invariant mass distribution 
of the $B^- \to \pi^- \mu^+ \mu^-$ decay \cite{Hambrock:2015wka}
\footnote{We denote explicitly by $C^{\rm SM}_{9,10}$ the SM Wilson coefficients,
$C^{\rm SM}_{9} (m_b) \approx - C^{\rm SM}_{10} (m_b) \approx 4.1 $}

\begin{eqnarray}
\label{eq:dBdqsq-NP}
\frac{d {\rm BR}^{\rm NP} (B^- \to \pi^- \mu^+ \mu^-)}{d q^2} = \tau_{B^-}
\frac{G_F^2 \alpha_{\rm em}^2 |V_{tb} V_{td}^*|^2}{1536 \pi^5 m_B^3} |f^+_{B\pi}
(q^2)|^2 \lambda^{3/2}(m_B^2,m_\pi^2,q^2)
\nonumber\\
\times \Bigg\{\left| C_9^{\rm SM} + C_9^{\rm NP} + \Delta C_9^{B\pi}(q^2) + 
\frac{2 m_b}{m_B + m_\pi} C_7^{\rm SM} \, \frac{f^T_{B\pi} (q^2)}
{f^+_{B\pi}(q^2)}\right|^2
+ \left| C_{10}^{\rm SM} + C_{10}^{\rm NP} \right|^2 \Bigg \} \,,
\end{eqnarray}
where the following notations are introduced:
\begin{equation}
C^{\rm NP}_9 \equiv C_9^{\mu} + C_9^{\prime \mu}, \qquad 
C^{\rm NP}_{10} \equiv C_{10}^{\mu} + C_{10}^{\prime \mu}.
\label{eq:C9NP-C10NP-def}
\end{equation}
In Eq.~(\ref{eq:dBdqsq-NP}), $f_{B\pi}^+ (q^2)$ and $f_{B\pi}^T (q^2)$ are the vector and tensor
$B \to \pi$ transition form factors, respectively, $\lambda (m_B^2, m_\pi^2, q^2)$ 
is the K\"allen function, and $\Delta C_9^{B \pi} (q^2)$ denotes the $q^2$-dependent 
effective Wilson coefficient accumulating contributions from the non-local hadronic amplitudes.
The definitions of above mentioned quantites and functions are given in Ref.~\cite{Hambrock:2015wka}.
The non-perturbative input include the form factors and non-local hadronic amplitudes.
The former were determined using the Light-Cone Sum Rules (LCSR) method 
while the latter were obtained using combination of the QCD factorisation and LCSR methods 
with hadronic dispersion relations 
(see Refs. \cite{Hambrock:2015wka, Khodjamirian:2017fxg} for details).
In the numerical analysis we use the same input as in Ref.~\cite{Khodjamirian:2017fxg}.
We note that due to parity conservation in QCD the hadronic matrix element
\begin{equation}
\langle \pi (p) | \bar d \gamma_\mu \gamma_5 b | B(p+q) \rangle = 0
\end{equation}
vanishes and therefore it is not possible to resolve the contributions from 
left- and right-handed quark operators in the $B \to \pi \mu^+ \mu^-$ decays.

\begin{figure}[t]\centering
\includegraphics[scale=0.37]{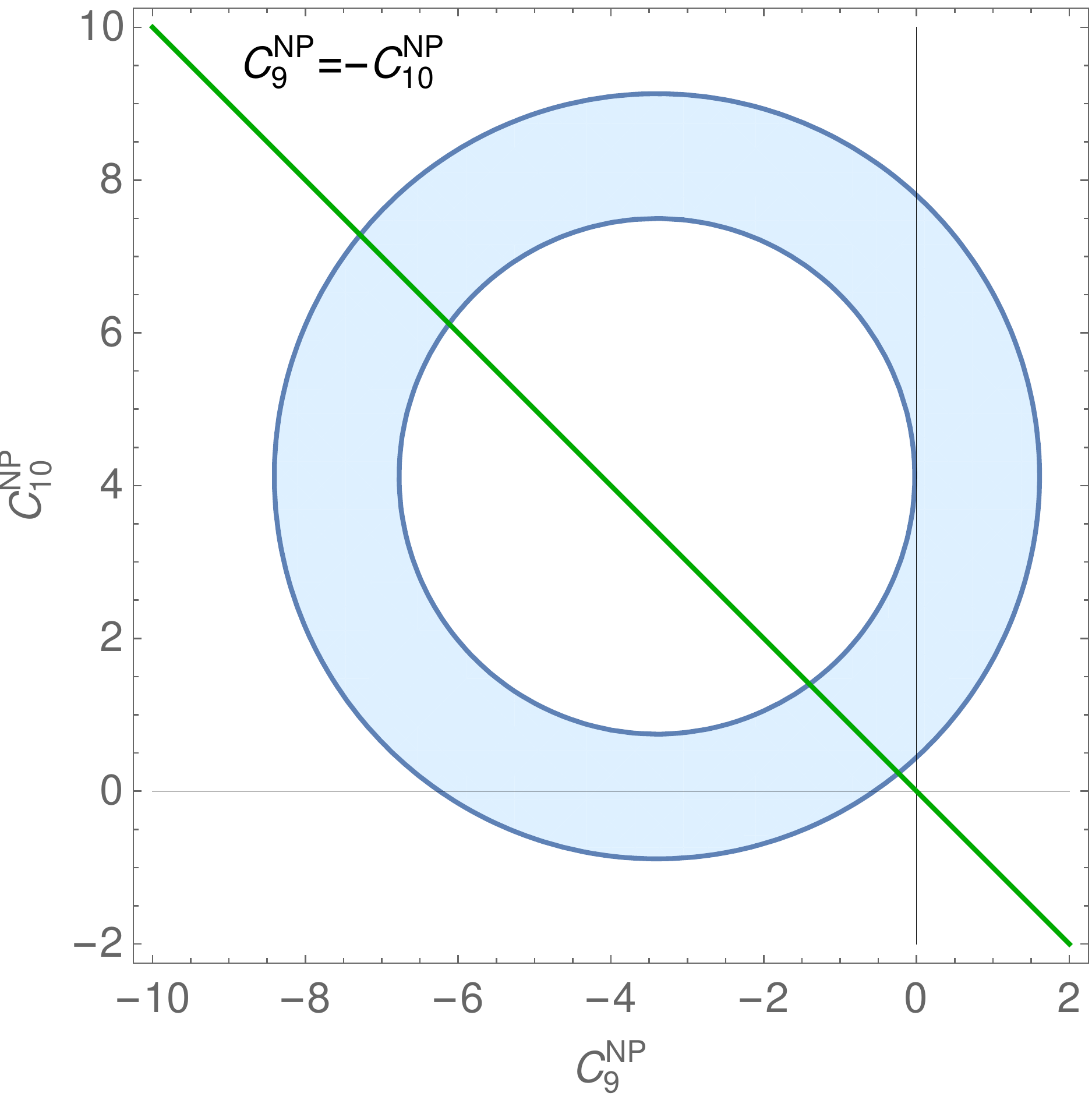}
\hfill
\includegraphics[scale=0.37]{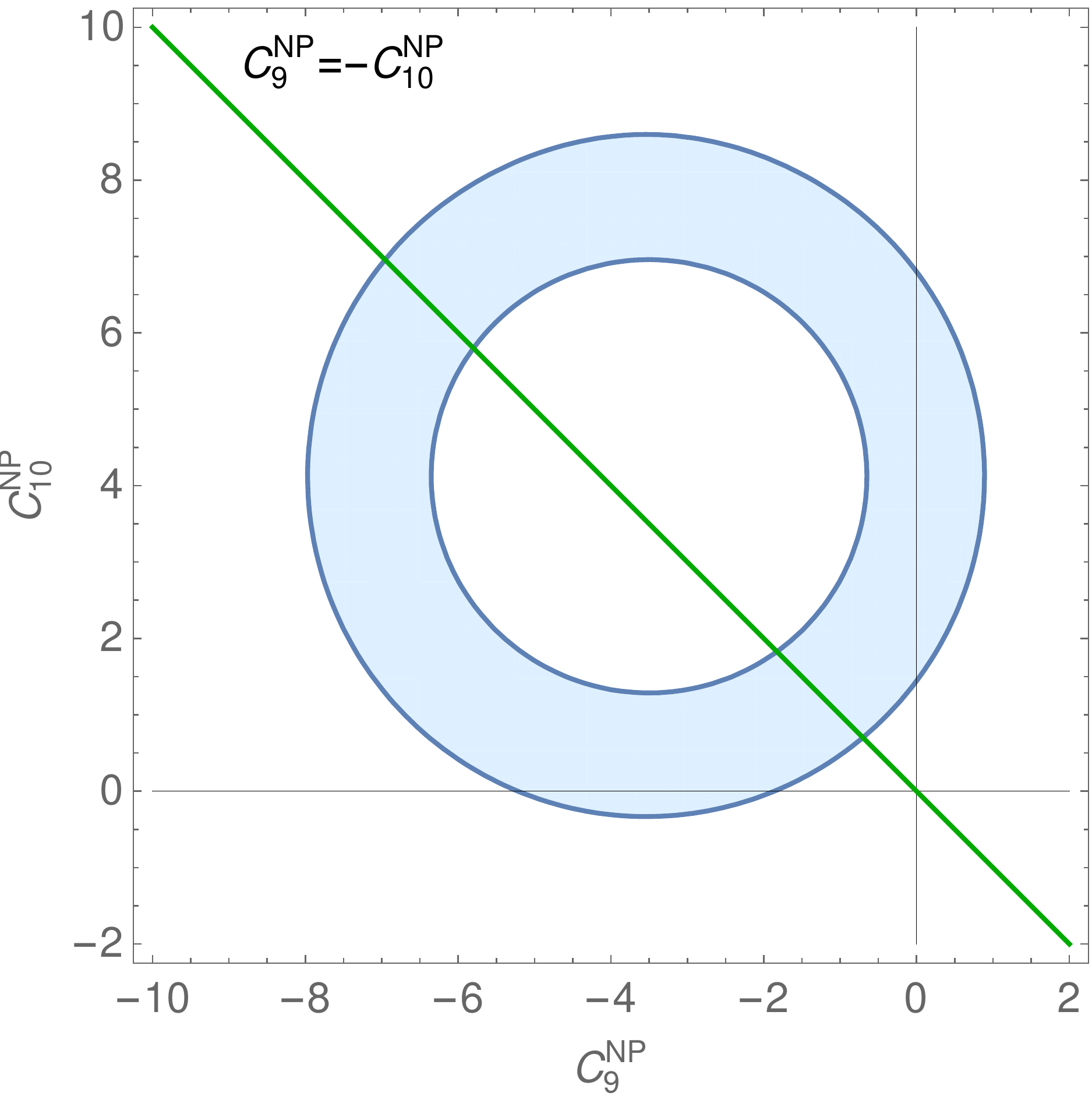}
\caption{Estimated $1 \sigma$-regions of $C_9^{\rm NP}$ and $C_{10}^{\rm NP}$
from the fit using one bin (left plot) and three bins (right plot). 
The green lines correspond to the scenario $C_{9}^{\rm NP} = - C_{10}^{\rm NP}$}
\label{fig:C9-C10-NP}
\end{figure}

We define the CP-averaged bin of the dilepton invariant mass distribution as
\begin{equation}
{\cal B} [q_1^2,q_2^2] \equiv
\frac{1}{2} 
\frac{1}{q_2^2-q_1^2}\int\limits_{q_1^2}^{q_2^2} \! dq^2 \!
\left[ \frac{d {\rm BR}  (B^- \to \pi^- \ell^+ \ell^-)}{d q^2}\,
+ \frac{d {\rm BR} (B^+ \to \pi^+ \ell^+ \ell^-)}{d q^2} \right].
\label{eq:bin}
\end{equation}
The SM prediction for this observable in the bin $[1 - 6]\, {\rm GeV}^2$
presented in Table~5 of~Ref.~\cite{Khodjamirian:2017fxg} is about 1.3$\sigma$
above the corresponding experimental measurement by the LHCb collaboration
\cite{Aaij:2015nea}. 
Experimental values of ${\cal B} [q_1^2, q_2^2]$ for three bins 
$[2 - 4]\, {\rm GeV}^2$, $[4 - 6]\, {\rm GeV}^2$ and $[6 - 8] \, {\rm GeV}^2$
\cite{Aaij:2015nea} are also not directly overlapping with the SM predictions.
Using these experimental data we perform a fit of the 
NP coefficients $C_9^{\rm NP}$ and $C_{10}^{\rm NP}$ assuming them to be real.
Note that we do not include the bin $[0.1 - 2]\, {\rm GeV}^2$
near the $\rho$- and $\omega$-resonances due to large hadronic uncertainties arising
in their theoretical description.
The fit is performed by using the method of least squares
introducing the $\chi^2$~function
\begin{equation}
\chi^2 = \sum_{i = 1}^{N_b} 
\frac{\left({\cal B}_i^{\rm \, NP} - {\cal B}_i^{\rm \, exp} \right)^2}{\sigma_i^2},
\label{eq:chi-sq-func}
\end{equation} 
where $N_b$ is the number of bins, ${\cal B}_i^{\rm \, NP}$ denotes 
the theoretical expression for the bin of the dimuon invariant mass distribution 
depending on the NP Wilson coefficients, 
and ${\cal B}_i^{\rm \, exp}$~is the corresponding experimental measurement.
Both theoretical and experimental uncertainties are assumed to be Gaussian distributed,
no correlations between experimental values of the bins are quoted 
in Ref.~\cite{Aaij:2015nea}. The theory predictions for the bins are in general correlated
between each other but we neglect these effects in our analysis, 
since the uncertainty of fit is mostly dominated by the experimental errors.
In the future, the analysis can be improved by including the correlations between bins
when more accurate data will be available.  
The standard deviation $\sigma_i$ in Eq.~(\ref{eq:chi-sq-func}) includes both experimental
and theoretical uncertainties in quadrature. Note that
theory uncertainties are determined only for vanishing NP Wilson coefficients.
In our analysis we consider the following scenarios: 
(1) only $C_9^{\rm NP}$; (2) only $C_{10}^{\rm NP}$; 
(3) both $C_9^{\rm NP}$ and $C_{10}^{\rm NP}$ as independent from each other; 
and (4) $C_9^{\rm NP} = - C_{10}^{\rm NP}$.
The results obtained are presented in Table~\ref{Tab:res-C9-C10-NP}
and Fig.~\ref{fig:C9-C10-NP}.

\begin{table}[t]\centering
\begin{tabular}{|l|c|c|c|}
\hline
\multicolumn{4}{|c|}{One bin} \\
\hline
Scenario & Best-fit values & 1$\sigma$ interval & 
Pull \\
\hline
$C_9^{\rm NP}$ only						
& $-2.3; \, -4.5 $ & $[-6.2, -0.6]$
& 1.5 \\
$C_{10}^{\rm NP}$ only 
& $+1.5; \, +6.7 $ & $[+0.4, +7.8]$ 
& 1.5 \\
$C_{9}^{\rm NP} = - C_{10}^{\rm NP}$	
& $-0.8, -6.7 $ & $ [-1.4, -0.2] \cup [-7.3, -6.1]$ 
& 1.5 \\
\hline
${\rm both} \; C_{9}^{\rm NP} \; {\rm and} \; C_{10}^{\rm NP}$	
& \multicolumn{3}{|c|}{see Fig.~\ref{fig:C9-C10-NP} } \\
\hline
\end{tabular}
\\[5mm]

\begin{tabular}{|l|c|c|c|}
\hline
\multicolumn{4}{|c|}{Three bins} \\
\hline
Scenario & Best-fit value(s) & 1$\sigma$ interval & 
Pull \\
\hline
$C_9^{\rm NP}$ only						
& $-3.6$ & $[-5.2, -1.9]$ 
& 2.6 \\
$C_{10}^{\rm NP}$ only 
& $+2.8; \, +5.4\, $ & $[+1.4, +6.8]$ 
& 2.7 \\
$C_{9}^{\rm NP} = - C_{10}^{\rm NP}$	
& $-1.2; \, -6.4 $ & $ [-1.8, -0.7] \cup [-7.0, -5.8]$ 
& 2.7 \\
\hline
${\rm both} \; C_{9}^{\rm NP} \; {\rm and} \; C_{10}^{\rm NP}$			
& \multicolumn{3}{|c|}{see Fig.~\ref{fig:C9-C10-NP} } \\
\hline
\end{tabular}
\caption{Estimated $1 \sigma$ ranges of the NP 
coefficients $C_9^{\rm NP}$ and $C_{10}^{\rm NP}$ in different scenarios.
"One bin" refers to the bin $[1 - 6]\, {\rm GeV}^2$, 
and "three bins" includes $[2 - 4]\, {\rm GeV}^2$, $[4 - 6]\, {\rm GeV}^2$ 
and $[6 - 8] \, {\rm GeV}^2$. Pull is defined as a square root of the difference
of $\chi^2$ values between the best-fit and SM points: 
pull = $\sqrt{\chi_{\rm SM}^2 - \chi_{\rm min}^2}$}
\label{Tab:res-C9-C10-NP}
\end{table}

\begin{figure}[t]\centering
\includegraphics[scale=0.45]{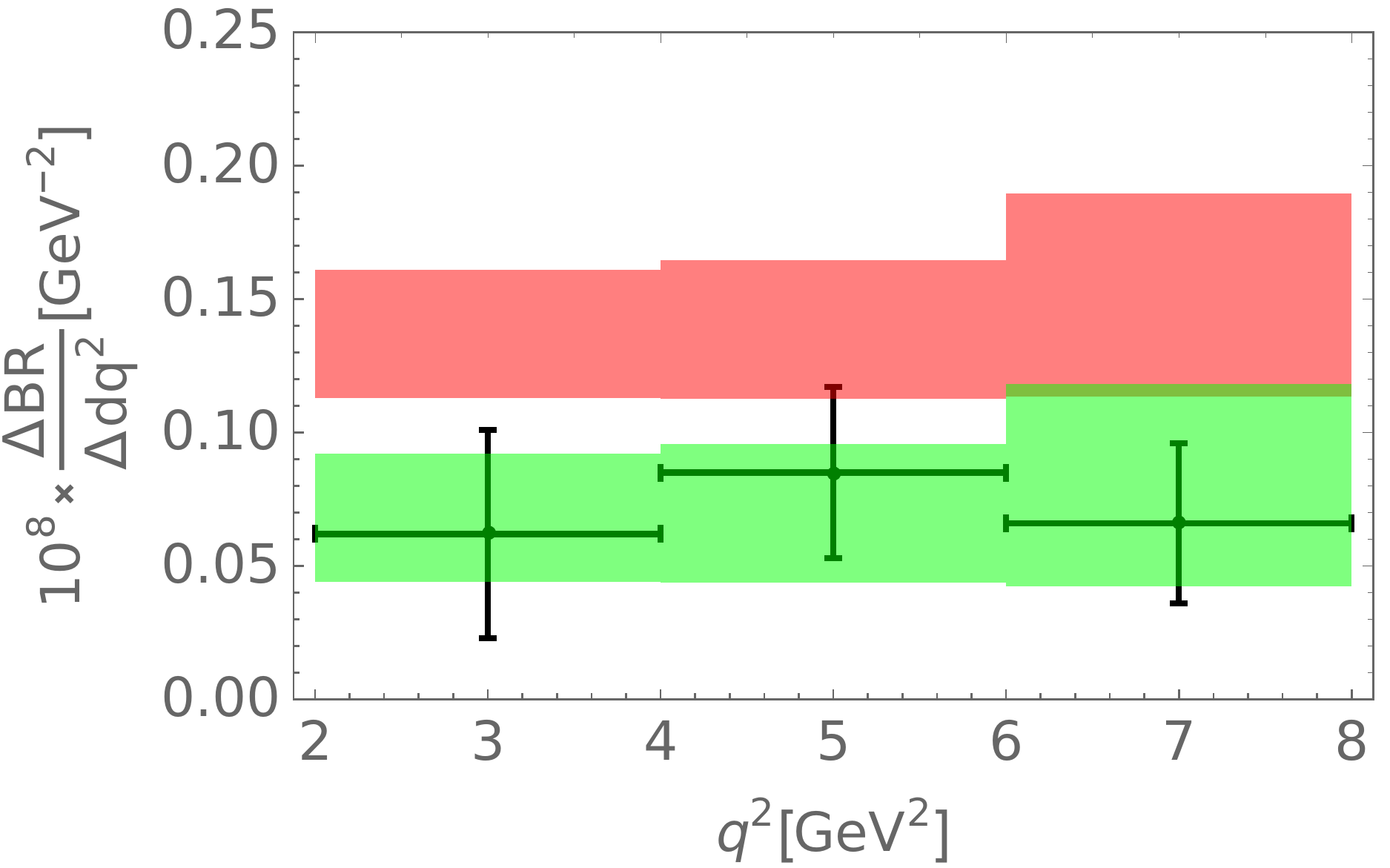}
\caption{Theoretical predictions for differential branching fraction of the 
$B^\pm \to \pi^\pm \mu^+ \mu^-$ decays in comparison with the data (black) by the LHCb collaboration
\cite{Aaij:2015nea}. The red bands correspond to the SM prediction \cite{Khodjamirian:2017fxg}
and green ones indicate the NP result (for the solution in Eq.~(\ref{eq:C9=-C10-interval}))
}
\label{fig:dBR-SM-data-NP}
\end{figure}

Let us make several comments on these results. First, we note 
that rather broad intervals are still allowed for separately $C_9^{\rm NP}$ 
and $C_{10}^{\rm NP}$, this is mainly due to large experimental errors. 
Second, considering scenario with $C_9^{\rm NP} = - C_{10}^{\rm NP}$ yields 
two separate solutions $C_9^{\rm NP} = - C_{10}^{\rm NP} \sim -6$ and
$C_9^{\rm NP} = - C_{10}^{\rm NP} \sim -1$ 
(see Fig.~\ref{fig:C9-C10-NP} and Table~\ref{Tab:res-C9-C10-NP}) 
where the latter one is quite similar\,\footnote{Note, that "similar" in this context 
just refers to the similar values of NP Wilson coefficients in $b \to d$ 
and $b \to s$ transitions, respectively,
as it follows from the normalisation of corresponding NP Lagrangian. 
In fact, the NP effects in $b \to d$ are not of similar size as in $b \to s$ 
and are suppressed by the ratio $|V_{td}/V_{ts}|$, for instance as a consequence 
of minimally broken $U(2)$-flavour symmetry, see e.g. Ref.~\cite{Fuentes-Martin:2019mun}}
to the $b \to s \ell^+ \ell^-$ case.
Hereafter we focus on consideration the following solution
\begin{equation}
C_9^{\rm NP} = - C_{10}^{\rm NP} \simeq -1.2 \pm 0.6.
\label{eq:C9=-C10-interval}
\end{equation}
It is interesting to notice, that from the global fit of data on the $b \to s \ell^+ \ell^-$
observables one also gets quite similar estimates (updated after Moriond 2019):
\begin{eqnarray}
C_{9, \, bs}^{\mu} = - C_{10, \, bs}^{\mu} = - 0.46 \pm 0.10, & &
\mbox{\cite{Alguero:2019ptt}} \\
C_{9, \, bs}^{\mu} = - C_{10, \, bs}^{\mu} = - 0.41 \pm 0.10, & & 
\mbox{\cite{Arbey:2019duh}} \\
C_{9, \, bs}^{\mu} = - C_{10, \, bs}^{\mu} = - 0.53 \pm 0.08, & & 
\mbox{\cite{Aebischer:2019mlg}}
\end{eqnarray}
assuming the LFU violation in $\mu$-$e$ sector.
A comparison of the SM prediction \cite{Khodjamirian:2017fxg}, 
LHCb data \cite{Aaij:2015nea} and the NP result (in the scenario of Eq.~(\ref{eq:C9=-C10-interval})) 
for the binned differential branching fraction of the $B^\pm \to \pi^\pm \mu^+ \mu^-$ decays 
is presented in Fig.~\ref{fig:dBR-SM-data-NP}. 
We again emphasize that our estimate in Eq.~(\ref{eq:C9=-C10-interval}) still allows
for both left- and right-handed quark operators ${\cal O}_{9,10}^\mu$ 
and ${\cal O}_{9,10}^{\prime \mu}$ as one can see from Eq.~(\ref{eq:C9NP-C10NP-def}).
This ambiguity can be resolved by considering the leptonic $B^0 \to \mu^+ \mu^-$
decay which is sensitive to another combination of the effective operators.
This question is discussed in the next section.


\section{Impact on the \boldmath $B^0 \to \mu^+ \mu^-$ decay}

There are several experimental analyses of the $B^0 \to \mu^+ \mu^-$ decay
performed by the ATLAS, CMS and LHCb collaborations \cite{Chatrchyan:2013bka, CMS:2014xfa, Aaboud:2016ire, Aaboud:2018mst, CMS:2019qnb, Aaij:2017vad}.
No significant evidence of $B^0 \to \mu^+ \mu^-$ decay was found so far,
and only upper limits are set up. The most recent bounds are
\begin{eqnarray}
{\rm BR} (B^0 \to \mu^+ \mu^-) < 2.1 \times 10^{-10}, \qquad 95\% \, {\rm CL},
& & \mbox{\cite{Aaboud:2018mst}} \quad {\rm (ATLAS)}
\label{eq:BrBmumu-ATLAS-2018} \\
{\rm BR} (B^0 \to \mu^+ \mu^-) < 3.6 \times 10^{-10}, \qquad 95\% \, {\rm CL},
& & \mbox{\cite{CMS:2019qnb}} \quad {\rm (CMS)}
\label{eq:BrBmumu-CMS-2019} \\
{\rm BR} (B^0 \to \mu^+ \mu^-) < 3.4 \times 10^{-10}, \qquad 95\% \, {\rm CL}.
& & \mbox{\cite{Aaij:2017vad}} \quad {\rm (LHCb)}
\label{eq:BrBmumu-LHCb-2017} 
\end{eqnarray}
The Particle Data Group quotes the average value \cite{PDG2018} (online update)
based on combination of the results in Refs.~\cite{Aaboud:2018mst, Aaij:2017vad, CMS:2014xfa} 
\begin{equation}
{\rm BR} (B^0 \to \mu^+ \mu^-) = (1.4^{+1.6}_{-1.4}) \times 10^{-10} 
\label{eq:BrBmumu-average}
\end{equation}
that is consistent with zero.

The SM prediction for the $B^0 \to \mu^+ \mu^-$ decay width is known up to ${\cal O} (\alpha_{em})$ 
and ${\cal O} (\alpha_s^2)$ corrections \cite{Bobeth:2013uxa, Beneke:2017vpq, Beneke:2019slt}. 
Using values for the decay constant from Lattice QCD \cite{Aoki:2019cca} the most recent SM value 
for the $B^0 \to \mu^+ \mu^-$ decay branching fraction is obtained in Ref.~\cite{Beneke:2019slt}
\begin{equation}
{\rm BR}^{\rm SM} (B^0 \to \mu^+ \mu^-) = (1.027 \pm 0.051) \times 10^{-10}.
\label{eq:BrBmumu-SM}
\end{equation}
This result is consistent with experimental upper limits in Eqs.~(\ref{eq:BrBmumu-ATLAS-2018}),
(\ref{eq:BrBmumu-CMS-2019}), (\ref{eq:BrBmumu-LHCb-2017}) 
as well as with the average in Eq.~(\ref{eq:BrBmumu-average}).
The NP Lagrangian (\ref{eq:Heff-NP}) leads to a modification 
of the branching fraction of the $B^0 \to \mu^+ \mu^-$ decay also induced at quark level 
by the $b \to d \mu^+ \mu^-$ transition.
The modified expression for the $B^0 \to \mu^+ \mu^-$ branching fraction reads:
\begin{equation}
{\rm BR}^{\rm NP} (B^0 \to \mu^+ \mu^-) = \tau_{B^0}
\frac{G_F^2 \alpha_{\rm em}^2 |V_{tb}^* V_{td}|^2}{16 \pi^3} 
m_{B^0} f_{B}^2 m_\mu^2 \sqrt{1 - \frac{4 m_\mu^2}{m_{B^0}^2}} 
\left|C_{10}^{\rm SM} + C_{10}^\mu - C_{10}^{\prime \mu} \right|^2.
\label{eq:Bd-to-mu-mu-NP}
\end{equation}
As one can see from Eq.~(\ref{eq:Bd-to-mu-mu-NP}), the left- and right-handed quark
current operators ${\cal O}_{10}$ and ${\cal O}_{10}^\prime$ give opposite sign
contributions to the $B^0 \to \mu^+ \mu^-$ branching fraction.
Keeping in mind the relation (\ref{eq:C9NP-C10NP-def}) and considering 
two cases with left- and right-handed operators separately, using the value in Eq.~(\ref{eq:C9=-C10-interval})
we get the following NP estimates for the~$B^0\to\mu^+\mu^-$~decay  branching fraction, respectively:
\begin{eqnarray}
{\rm BR}^{\rm NP} (B^0 \to \mu^+ \mu^-) & \simeq & (0.6 \pm 0.2) \times 10^{-10},  
\qquad \mbox{if  } C_{10}^{\rm NP} = C_{10},
\label{BR-B-to-mu-mu-NP-left} \\
{\rm BR}^{\rm NP} (B^0 \to \mu^+ \mu^-) & \simeq & (1.8 \pm 0.4) \times 10^{-10},  
\qquad \mbox{if  } C_{10}^{\rm NP} = C_{10}^\prime.
\label{BR-B-to-mu-mu-NP-right}
\end{eqnarray}
Both values above are consistent with the experimental bounds (\ref{eq:BrBmumu-ATLAS-2018}),
(\ref{eq:BrBmumu-CMS-2019}), (\ref{eq:BrBmumu-LHCb-2017})
while the value in Eq.~(\ref{BR-B-to-mu-mu-NP-right}) is quite close to the upper limit 
by ATLAS collaboration (\ref{eq:BrBmumu-ATLAS-2018}).
Therefore, currently we are not able to make an unambiguous conclusion regarding 
preference of left- or right-handed quark currents in $b \to d \mu^+ \mu^-$ transition. 
Nevertheless, future more precise data on the~$B^0 \to \mu^+ \mu^-$ decay would clarify this situation.

\section{Impact on \boldmath $B^0 - \bar B^0$ mixing}
We would like to emphasize that in general the model-independent Lagrangian
(\ref{eq:Heff-NP}) does not necessarily give a sizeable impact on 
$B^0 - \bar B^0$ mixing. Indeed, the NP operators in the form (\ref{eq:NP-operators})
can give contribution to the mixing via muonic loops that are suppressed
compared to the tree-level contribution of the SM dimension-6 four-quark operator $Q_1~=~\bar d \gamma_\mu (1 - \gamma_5) b \times \bar d \gamma^\mu (1 - \gamma_5) b$.
However, depending on a specific model, the $B^0 - \bar B^0$ mixing might be strongly 
affected by NP in $b \to d$ transition.

The mass difference of the mass eigenstates in $B^0 - \bar B^0$ system 
is given by (see e.g. Refs.~\cite{DiLuzio:2017fdq}, \cite{Artuso:2015swg}):
\begin{equation}
\Delta M_d = 2 | M_{12}^d | = \frac{G_F^2}{6 \pi^2} |V_{tb} V_{td}^*|^2 m_W^2
S_0 (x_t) \hat \eta_B m_{B} f_{B}^2 B,
\label{eq:Delta-Md}
\end{equation}
where $S (x_t) \, (x_t = m_t^2/m_W^2)$ is the Inami-Lim function \cite{Inami:1980fz}, 
$\hat \eta_B$ encodes perturbative QCD corrections \cite{Buras:1990fn}, 
and $B$ denotes the Bag parameter characterising 
the matrix element of the dimension-6 operator $Q_1$.
Note that due to parity conservation of QCD the matrix element of 
the operator with right-handed currents $Q_1^\prime = 
\bar d \gamma_\mu (1 + \gamma_5) b \times \bar d \gamma^\mu (1 + \gamma_5) b$
is described by the same Bag parameter $B^\prime = B$.

The mass difference $\Delta M_d$ is measured very precisely, 
the value quoted by HFLAV in 2019 \cite{Amhis:2019ckw}
\begin{equation}
\Delta M_d^{\rm exp} = (0.5064 \pm 0.0019) \, {\rm ps}^{-1}
\label{eq:Delta-Md-exp}
\end{equation}
is in agreement with the average value \cite{DiLuzio:2019jyq} obtained using a combination of 
HQET Sum Rule \cite{Grozin:2016uqy, Grozin:2017uto, Grozin:2018wtg, Kirk:2017juj, King:2019lal} 
and Lattice QCD results \cite{Dowdall:2019bea, Boyle:2018knm, Aoki:2019cca}:
\begin{eqnarray}
\Delta M_d^{\rm average} = (0.533^{+0.022}_{-0.036}) \, {\rm ps}^{-1}.
\label{eq:Delta-Md-average}
\end{eqnarray}

\begin{figure}[t]\centering
\includegraphics[scale=0.27]{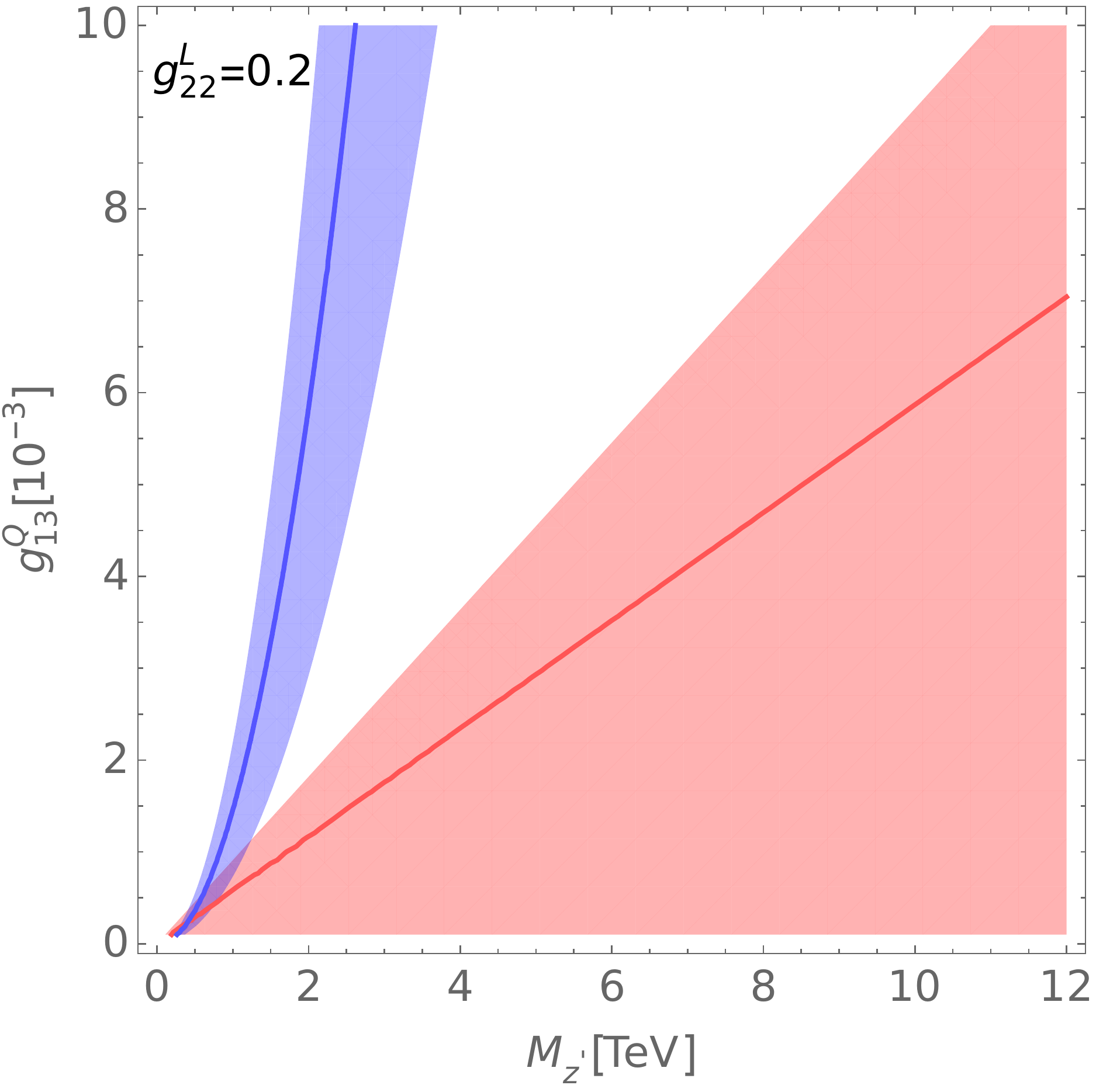}
\includegraphics[scale=0.27]{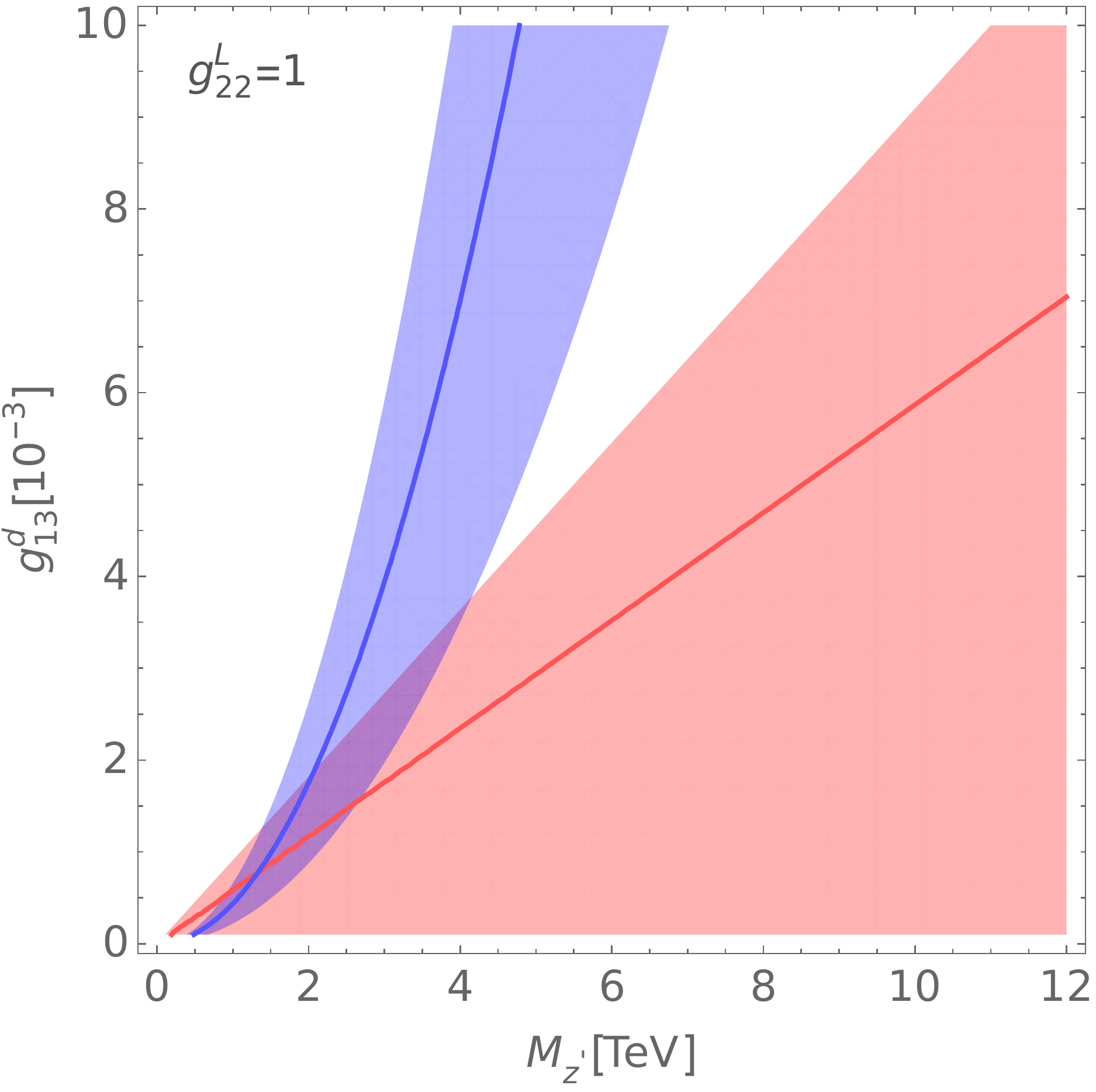}
\includegraphics[scale=0.27]{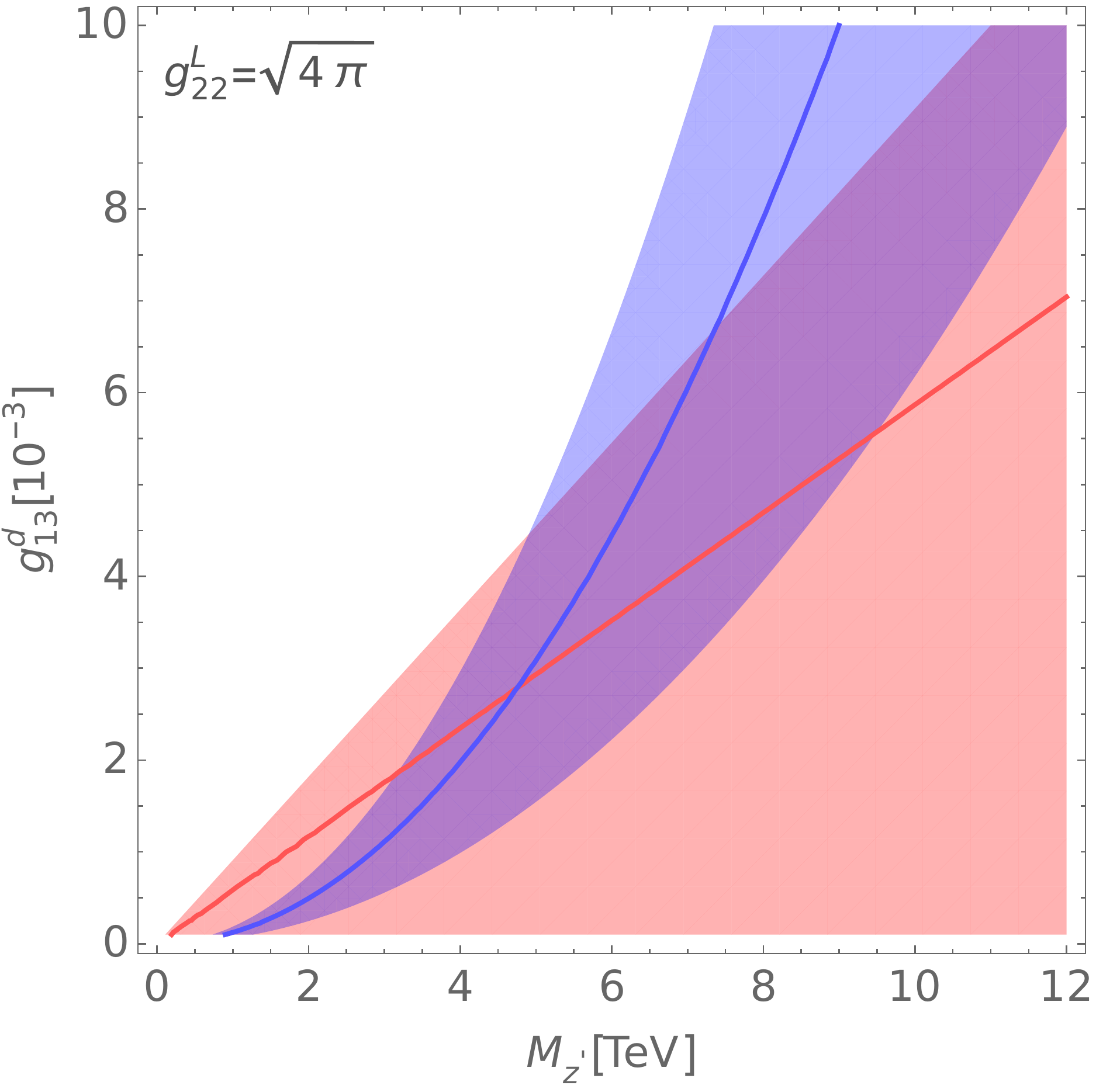}
\caption{Bounds from $B^0 - \bar B^0$ mixing on the coupling $g_{13}^Q$ 
and $m_{Z^\prime}$ for fixed $g_{22}^L = 0.2$ (left), $g_{22}^L = 1$ (middle)
and $g_{22}^L = \sqrt {4 \pi}$ (right). 
The red area corresponds to preferred $1 \sigma$~region from $\Delta M_d$ 
and the blue one to $1 \sigma$~region from $B^\pm \to \pi^\pm \mu^+ \mu^-$ decays.
}
\label{fig:g13-mZp}
\end{figure}

As an example we will investigate a simplified NP model with $Z^\prime$-boson that
couples with left-handed quarks and leptons in order to find a prefered range 
of NP parameters consistent with current $b \to d \mu^+ \mu^-$ data 
and $B^0 - \bar B^0$ mixng.
In the framework of this model we consider the following interaction Lagrangian \cite{DiLuzio:2019jyq}:
\begin{equation}
{\cal L}_{Z^\prime} = 
\left[g_{ij}^Q (\bar d_L^{\, i} \gamma^\mu d_L^j) + 
g_{ij}^L (\bar \ell_L^i \gamma^\mu \ell^j_L) \right] Z_\mu^\prime + {\rm h.c.}.
\label{eq:L-Zp}
\end{equation}
Integrating out the heavy $Z^\prime$-boson yields the following effective Lagrangian
\begin{equation}
{\cal L}_{Z^\prime}^{\rm eff} = 
-\frac{1}{2 m_{Z^\prime}^2} \left[g_{ij}^Q (\bar d_L^{\, i} \gamma^\mu d_L^j) + 
g_{ij}^L (\bar \ell_L^i \gamma^\mu \ell^j_L) \right]^2 + {\rm h.c.}.
\label{eq:L-Zp-eff}
\end{equation}
In the above, we hereafter consider only terms relevant 
for the $b \to d \mu^+ \mu^-$ transition and $B^0 - \bar B^0$-mixing:
\begin{equation}
{\cal L}_{Z^\prime}^{\rm eff} = -\frac{1}{2 m_{Z^\prime}^2} 
\left[\left(g_{13}^Q \right)^2 
(\bar d_L \gamma^\mu b_L)(\bar d_L \gamma_\mu b_L) + 2 g_{13}^Q \, g_{22}^L \,
(\bar d_L \gamma^\mu b_L) (\bar \mu_L \gamma_\mu \mu_L) \right] + \ldots .
\label{eq:L-Zp-eff-bd}
\end{equation}
Parametrising NP effects in mass difference $\Delta M_d$ as
\begin{equation}
\frac{\Delta M_d^{\rm exp}}{\Delta M_d^{\rm SM}} = 
\left|1 + \frac{C_{bd}^{LL}}{R}\right|,
\label{eq:Delta-Md-ratio}
\end{equation}
where  $R =\sqrt 2 G_F m_W^2 S_0 (x_t) \hat\eta_B/(16 \pi^2)
\approx 1.34 \times 10^{-3}$, 
and taking into account the expression for the effective NP Lagrangians
(\ref{eq:Heff-NP}) and (\ref{eq:L-Zp-eff-bd}), 
one gets the following relation between the NP coefficients
and parameters of the simplified $Z^\prime$ model  
\cite{DiLuzio:2017fdq, DiLuzio:2019jyq}:
\begin{equation}
C_9^\mu = - C_{10}^\mu = 
- \frac{\sqrt 2 \pi}{2 G_F m_{Z^\prime}^2 \alpha_{\rm em}}
\left(\frac{g_{13}^Q \, g_{22}^L}{V_{tb} V_{td}^*} \right), 
\label{eq:C9-Zp-model}
\end{equation}
\begin{equation}
C_{bd}^{LL} = \frac{\eta(m_{Z^\prime})}{4 \sqrt 2 G_F m_{Z^\prime}^2} 
\left(\frac{g_{13}^Q}{V_{tb} V_{td}^*} \right)^{\! 2},
\label{eq:Cbd-Zp-model}
\end{equation}
where $\eta (m_{Z^\prime}) = (\alpha_s (m_{Z^\prime}) / \alpha_s (m_b))^{6/23}$ 
accounts for running from the $m_{Z^\prime}$ scale down to the $b$-quark mass scale.
Assuming that $C_9^{\rm NP} = - C_{10}^{\rm NP}$ in Eq.~(\ref{eq:C9=-C10-interval}) is 
given by left-handed quark currents and the coupling $g_{13}^Q$ is real,
taking into account Eqs.~(\ref{eq:Delta-Md-ratio}), (\ref{eq:C9-Zp-model}), 
and (\ref{eq:Cbd-Zp-model})  
and using the experimental (\ref{eq:Delta-Md-exp})
and average (\ref{eq:Delta-Md-average}) values for $\Delta M_d$ 
we get constraints on parameters $g_{13}^Q$ and $m_{Z^\prime}$ presented in Fig.~\ref{fig:g13-mZp} 
(for three different reference values of $g_{22}^L = 0.2$, $g_{22}^L = 1$ and 
$g_{22}^{L} = \sqrt{4 \pi}$).
The red area corresponds to favored values from mixing and the blue one from the data 
on the $B^\pm \to \pi^\pm \mu^+ \mu^-$ decays, both at $1 \sigma$~level
(in the scenario given in Eq.~(\ref{eq:C9=-C10-interval})).
We notice that prefered area of $g_{13}^Q$ and $m_{Z^\prime}$
parameters shrinks with smaller values of $q_{22}^L$ 
as one can see from comparing the plots in Fig.~\ref{fig:g13-mZp} from right to left.

So, we arrive to the conclusion that a simplified model with $Z^\prime$-boson that 
couples with left-handed $b \to d$ quark current might potentially explain current data on 
$B^\pm \to \pi^\pm \mu^+ \mu^-$ and $B^0 \to \mu^+ \mu^-$ decays without spoiling $B^0$-mixing 
only for relatively large values of $g_{22}^L$.
On the other side, small coupling $g_{22}^L$ disfavor 
large values of $Z^\prime$ mass, as one can see from the left plot of Fig.~\ref{fig:g13-mZp}.
Curiously, more or less the same picture is found in the case of 
$b \to s$ transition, see e.g. Ref.~\cite{DiLuzio:2019jyq}.

\section{Conclusion and discussion}

In contrast to the well studied and measured $b \to c$ and $b \to s$ flavour transitions 
where several anomalies have been found, the $b \to d \ell^+ \ell^-$ processes are so far 
poorly investigated experimentally due to an additional suppression by CKM matrix elements.
Nevertheless, recent experimental data by the LHCb collaboration for the differential
$q^2$-distribution in $B^\pm \to \pi^\pm \mu^+ \mu^-$ decays 
deviate a bit more than 1$\sigma$ from the recent Standard Model prediction. 
Interestingly, this slight tension  points in the same direction as in $b \to s \mu^+ \mu^-$ decays.
In this work, we performed a model-independent fit and obtained $1 \sigma$~intervals for the 
NP Wilson coefficients $C_9^{\rm NP}$ and $C_{10}^{\rm NP}$ in different scenarios. 
Note that our results allow so far for both left- and -right-handed quark currents despite 
the latter is quite close to the  experimental bound on $B^0 \to \mu^+ \mu^-$ decay.
Considering a specific simplified model with a $Z^\prime$-boson 
that couples with left-handed fermions ($b - d$ and $\mu -\mu$ currents)
we found an $1 \sigma$ range of NP parameters (the couplings 
$g_{13}^Q$, $g_{22}^L$ and $Z^\prime$-boson mass $m_{Z^\prime}$) 
that is consistent with current experimental data on 
$B^\pm \to \pi^\pm \mu^+ \mu^-$ and $B^0 \to \mu^+ \mu^-$ decays 
and $B^0 - \bar B^0$ mixing.

To make more robust statements concerning New Physics in the $b \to d$ sector
more experimental data on semileptonic and leptonic $b \to d$ processes will be necessary, 
including
(1) a more precise measurement of $B \to \pi \mu^+ \mu^-$ decays,
(2) an upcoming measurement of $\bar B_s^0 \to K^{0*} \mu^+ \mu^-$,
(3) a first measurement of $B \to \rho \mu^+ \mu^-$,
(4) a more accurate measurement of $B^0 \to \mu^+ \mu^-$.
Additionally, measurements of the semileptonic $b \to d \ell^+ \ell^-$ 
processes with electrons or $\tau$-leptons will provide an additional 
test of the Lepton Flavour Universality in the SM.

\section*{Acknowledgement}

I am very grateful to Alexander Lenz, Alexander Khodjamirian and Martin Bauer 
for useful comments and discussion and for careful reading of the manuscript. 
I appreciate the motivating communication with Andreas Crivellin. 
This work was supported by STFC through the IPPP grant.

\section*{Appendix: Correlation between \boldmath 
$B \to K \ell^+ \ell^-$ and \\ $B \to \pi \ell^+ \ell^-$ decays}

We consider the ratio of the partially integrated branching fraction of 
$B^- \to K^- \ell^+ \ell^-$ and $B^- \to \pi^- \ell^+ \ell^-$ decays,
which can be written as \cite{Khodjamirian:2017fxg}
\begin{eqnarray}
\frac{{\cal B}(B^- \to \pi^- \ell^+\ell^-[q_1^2,q_2^2])}
{{\cal B}(B^- \to K^- \ell^+ \ell^-[q_1^2,q_2^2])} =
\left|\frac{V_{td}}{V_{ts}}\right|^2 
\frac{{\cal F}_{B\pi}[q_1^2,q_2^2]}{{\cal F}_{BK}[q_1^2,q_2^2]}
\Bigg\{ 1 + \kappa_d^2 \, 
\frac{{\cal D}_{B\pi}[q_1^2,q_2^2]}{{\cal F}_{B\pi}[q_1^2,q_2^2]} 
\nonumber \\
+ 2 \kappa_d \Biggl(\cos  \xi_d \, 
\frac{{\cal C}_{B\pi}[q_1^2,q_2^2]}{{\cal F}_{B\pi}[q_1^2,q_2^2]}
- \sin  \xi_d \, 
\frac{{\cal S}_{B\pi}[q_1^2,q_2^2]}{{\cal F}_{B\pi}[q_1^2,q_2^2]}
\Biggr)
\Bigg\}\, ,
\label{eq:Bpi-to-BK-ratio}
\end{eqnarray}
where $\kappa_d \, e^{i \xi_d} = (V_{ub} V^*_{ud})/(V_{tb} V^*_{td})$.
In the above expression, the CKM matrix elements are explicitly isolated and 
the quantities ${\cal F}_{B K}, {\cal F}_{B\pi}, {\cal D}_{B\pi}, {\cal C}_{B\pi}$ 
and ${\cal S}_{B\pi}$ are CKM independent and accumulate contributions from Wilson coefficients, 
form factors, non-local hadronic amplitudes and phase space integration.
Explicit expressions of the above quantities can be found in Ref.~\cite{Khodjamirian:2017fxg} 
and their numerical values for the bin $[1-6] \, {\rm GeV}^2$ are quoted in Table~4 in
Ref.~\cite{Khodjamirian:2017fxg} where no correlations were taken into account.
Nevertheless, the $B \to K$ and $B \to \pi$ form factors have been determined
using the LCSR method, and due to common input 
involved in both sum rules the $B \to K$ and $B \to \pi$ form factors 
are actually correlated to each other.
To fill this gap, we improve the numerical analysis by accounting
the correlation between both LCSRs for vector $B \to \pi$ and $B \to K$ form factors
and as a consequence we calculate the correlations between the quantities 
${\cal F}_{B K}$, ${\cal F}_{B\pi}$, ${\cal D}_{B\pi}$, ${\cal C}_{B\pi}$,
and ${\cal S}_{B\pi}$.
In the corresponding statistical simulation we use the same input 
as in Ref.~\cite{Khodjamirian:2017fxg}.
The resulting correlation matrix for the bin $[1-6] \, {\rm GeV}^2$ is  
\begin{equation}
\left(
\begin{tabular}{c|ccccc}
& ${\cal F}_{BK}$ & ${\cal F}_{B\pi}$ & ${\cal D}_{B\pi}$ & 
${\cal C}_{B\pi}$ & ${\cal S}_{B\pi}$ \\
\hline
${\cal F}_{BK}$     & 1     & 0.53  & 0.02  & 0.08  & -0.09 \\
${\cal F}_{B\pi}$   & 0.53  & 1     & 0.07  & 0.24  & -0.15 \\
${\cal D}_{B\pi}$   & 0.02  & 0.07  & 1     & 0.83  & -0.34 \\
${\cal C}_{B\pi}$   & 0.08  & 0.24  & 0.83  & 1     & 0.03 \\
${\cal S}_{B\pi}$   & -0.09 & -0.15 & -0.34 & 0.03  & 1 \\
\end{tabular}
\right).
\label{eq:Correl-matr-res}
\end{equation}
The above matrix represents an addition to our numerical result
presented in Table~4 in Ref.~\cite{Khodjamirian:2017fxg}
and to be used in any further analysis, e.g. in determination of  
CKM matrix elements from the observables in the $B^\pm \to \pi^\pm \ell^- \ell^-$ 
and $B^\pm \to K^\pm \ell^- \ell^-$ decays (see Ref.~\cite{Khodjamirian:2017fxg} 
for more details) or for testing some BSM scenarios 
where one needs to account for possible correlation by considering
both $b \to s$ and $b \to d$ transitions.

\bibliographystyle{hieeetr}
\bibliography{References}

\end{document}